\documentclass[12pt]{article}
\usepackage{amsmath}
\usepackage{graphicx,psfrag,epsf}
\usepackage{enumerate}
\usepackage[square,numbers]{natbib}
\usepackage{multirow}
\usepackage{xcolor}
\usepackage{booktabs}
\usepackage{subcaption}
\usepackage{DNstyle}
\usepackage{soul} 
\usepackage{pifont}

\usepackage{rotating}

\usepackage[normalem]{ulem}

\newcommand{\blind}{1} 

\newcommand*{\bD}{\mbox{\mathversion{bold}$D$}}

\newcommand*{\BFX}{\mbox{\mathversion{bold}$X$}}

\newcommand*{\BFyi}{\mbox{\mathversion{bold}$y$}_i}

\newcommand*{\BFYi}{\mbox{\mathversion{bold}$Y$}_{\hspace*{-0.025in}i}}
\newcommand*{\mP}{\mbox{P}}
\newcommand*{\mOR}{\mbox{OR}}

\newcommand*{\bzero}{\mbox{\mathversion{bold}$0$}}

\newcommand*{\bJ}{\mbox{\mathversion{bold}$J$}}

\newcommand*{\BFalpha}{\mbox{\mathversion{bold}$\alpha$}}
\newcommand*{\BFbeta}{\mbox{\mathversion{bold}$\beta$}}
\newcommand*{\BFeta}{\mbox{\mathversion{bold}$\eta$}}
\newcommand*{\BFtau}{\mbox{\mathversion{bold}$\tau$}}
\newcommand*{\BFxi}{\mbox{\mathversion{bold}$\xi$}}

\begin{document}

\def\spacingset#1{\renewcommand{\baselinestretch}%
{#1}\small\normalsize} \spacingset{1}

\if1\blind
{
 \title{\bf Modeling semi-competing risks data as a longitudinal bivariate process}
 \author{Daniel Nevo\thanks{
 The authors gratefully acknowledge funding from National Institutes of Health grant R-01 CA181360.}\hspace{.2cm}\\
	Department of Statistics and Operations Research,\\ Tel Aviv University, Tel Aviv 69978, Israel; \vspace{0.3cm} \\
	Deborah Blacker\\
	Department of Epidemiology, \\ Harvard T. H. Chan School of Public Health, Boston, MA 02115, USA\\
	 Department of Psychiatry, Massachusetts General Hospital, \\ Harvard Medical School,  Boston, MA 02115, USA;\vspace{0.3cm} \\
	 Eric B. Larson\\
	 Kaiser Permanente Washington Health Research Institute,\\ Seattle, WA, USA \vspace{0.3cm} \\
	  and \vspace{0.3cm}\\
 Sebastien Haneuse\\
 Department of Biostatistics, Harvard T.H. Chan School of Public Health,\\ Boston, MA 02115, USA}
 \maketitle
} \fi

\if0\blind
{
 \begin{center}
 {\LARGE\bf Modeling semi-competing risks data as a longitudinal bivariate process}
\end{center}
 \medskip
} \fi

\bigskip
\newpage
\begin{abstract}
The Adult Changes in Thought (ACT) study is a long-running prospective study of incident all-cause dementia and Alzheimer's disease (AD). As the cohort ages, death (a terminal event) is a prominent competing risk for AD (a non-terminal event), although the reverse is not the case. As such, analyses of data from ACT can be placed within the semi-competing risks framework. Central to semi-competing risks, and in contrast to standard competing risks, is that one can learn about the dependence structure between the two events. To-date, however, most methods for semi-competing risks treat dependence as a nuisance and not a potential source of new clinical knowledge. We propose a novel regression-based framework that views the two time-to-event outcomes through the lens of a longitudinal bivariate process on a partition of the time scale. A key innovation of the framework is that dependence is represented in two distinct forms, \textit{local} and \textit{global} dependence, both of which have intuitive clinical interpretations. Estimation and inference are performed via penalized maximum likelihood, and can accommodate right censoring, left truncation and time-varying covariates. The framework is used to investigate the role of gender and having $\ge$1 APOE-$\epsilon4$ allele on the joint risk of AD and death.
\end{abstract}

\noindent%
\vfill

\spacingset{1.45} 

\newpage
\section{Introduction}

Alzheimer's disease (AD) is a brain disorder characterized by progressive dementia that slowly destroys memory and cognitive function. In 2018, an estimated 5.7 Americans were living with AD~\citep{alzheimer2018}. First described in 1906, research over the last 40 years has identified numerous risk factors for AD, including: age, family history, the apolipoprotein E (APOE)-$\epsilon4$ allele, midlife obesity, midlife hypertension, and diabetes~\citep{baumgart2015summary}. Factors that exhibit a protective effect include education and physical activity~\citep{baumgart2015summary}.

Many of these factors are also strongly associated with mortality. This suggests that the outcomes of AD and mortality may be dependent within individuals and, furthermore, that this dependence may be influenced by a range of factors. One way of framing such dependence is through consideration of whether and how AD influences the risk of death. This could be achieved by, say, fitting a Cox model with death as the endpoint and AD as a time-varying covariate. While informative, this approach shifts attention away from AD as an outcome. Moreover, it precludes viewing AD and death as a multivariate outcome for which the components may exhibit covariation. This latter framing, we believe, may be of substantial interest to individuals who are alive and cognitively intact.

Practically, studies of risk factors for AD often focus on the timing of a diagnosis and thus use survival analysis methods. Typically, such analyses often treat death as a censoring mechanism in the observed data. An alternative is the semi-competing risks paradigm which would consider AD and mortality simultaneously. In general terms, semi-competing risks refers to the setting where interest lies in some non-terminal time-to-event outcome, the occurrence of which is subject to a terminal event~\citep{fine2001semi, varadhan2014semicompeting, haneuse2016semi}. Let $T_1$ and $T_2$ denote the time to the non-terminal and terminal events, respectively. Key to semi-competing risks is that one can potentially observe \textit{both} $T_1$ and $T_2$ on individual study units. As such, in contrast to the standard competing risks setting, there is partial information on the joint distribution of the non-terminal and terminal events in the semi-competing risks setting~\citep{tsiatis1975nonidentifiability, jazic2016beyond}. This, in turn, provides an opportunity to learn about the dependence structure between $T_1$ and $T_2$.

Beyond the usual challenges of time-to-event analyses (i.e. structuring covariate effects, handling functions of time, and accommodating various forms of censoring and truncation), key challenges that arise in the analysis of semi-competing risks data are: (i) respecting the terminal event as a competing risk; and, (ii) structuring dependence between $T_1$ and $T_2$. In the statistical literature, numerous frameworks for the analysis of semi-competing risks data have been proposed, including: methods grounded in causal inference~\citep{zhang2003estimation, egleston2006causal, tchetgen2014identification}; methods based on structuring dependence via a copula~\citep{fine2001semi, peng2007regression, hsieh2008regression, li2015quantile}; the use of illness-death models~\citep{xu2010statistical, lee2017accelerated, lee2015bayesian}; and, the recently-proposed cross-quantile residual ratio~\cite{yang2016new}. While additional review details are provided in Section A.1 of the Supplementary Materials, we note that these methods either: (i) view dependence as a statistical nuisance, and not a potential source of new clinical knowledge; or, (ii) focus on the role of the non-terminal event as a risk factor for the terminal event, thereby reframing the non-terminal event away from being an outcome of interest. As such, collectively, these methods fail to take advantage of the opportunity to learn about dependence between the two outcomes that semi-competing risks data provide.

In this paper we propose a novel regression-based framework for semi-competing risks data that simultaneously structures covariate effects on $T_1$ and on $T_2$, as well as on the dependence between the two events. A key innovation of the proposed framework is that dependence is represented in two distinct forms, termed the \textit{local} and \textit{global} dependence, both of which have intuitive clinical interpretations. Practically, smoothness in model components that are functions of time is facilitated through the use of B-splines, with estimation and inference via maximum penalized likelihood. Since the modeling framework is based upon probabilities conditional on survival, left truncation, right censoring and time-varying covariates may be accommodated in a straightforward manner. 

\section{The Adult Changes in Thought study}
\label{Sec:ACT}

The Adult Changes in Thought (ACT) study is an on-going community-based prospective study of incident all-cause dementia and AD among the elderly in western Washington state~\cite{kukull2002dementia}. Initiated in 1994, the goals of the study are to learn about how the brain ages and to identify risk factors for AD. In this paper, we consider data on $N$=4,367 ACT participants, who were enrolled between 1994-2015, and who were aged 65 years or older and cognitively intact at the time of enrollment. Table A.1 of Section A.2 in the Supplementary Materials provides a summary of key participant characteristics measured at study entry, including: age, gender, race, marital status, education, co-morbid depression, and APOE $\epsilon4$ carrier status.

\begin{figure}[h!]
\begin{center}
	\includegraphics[scale=0.55]{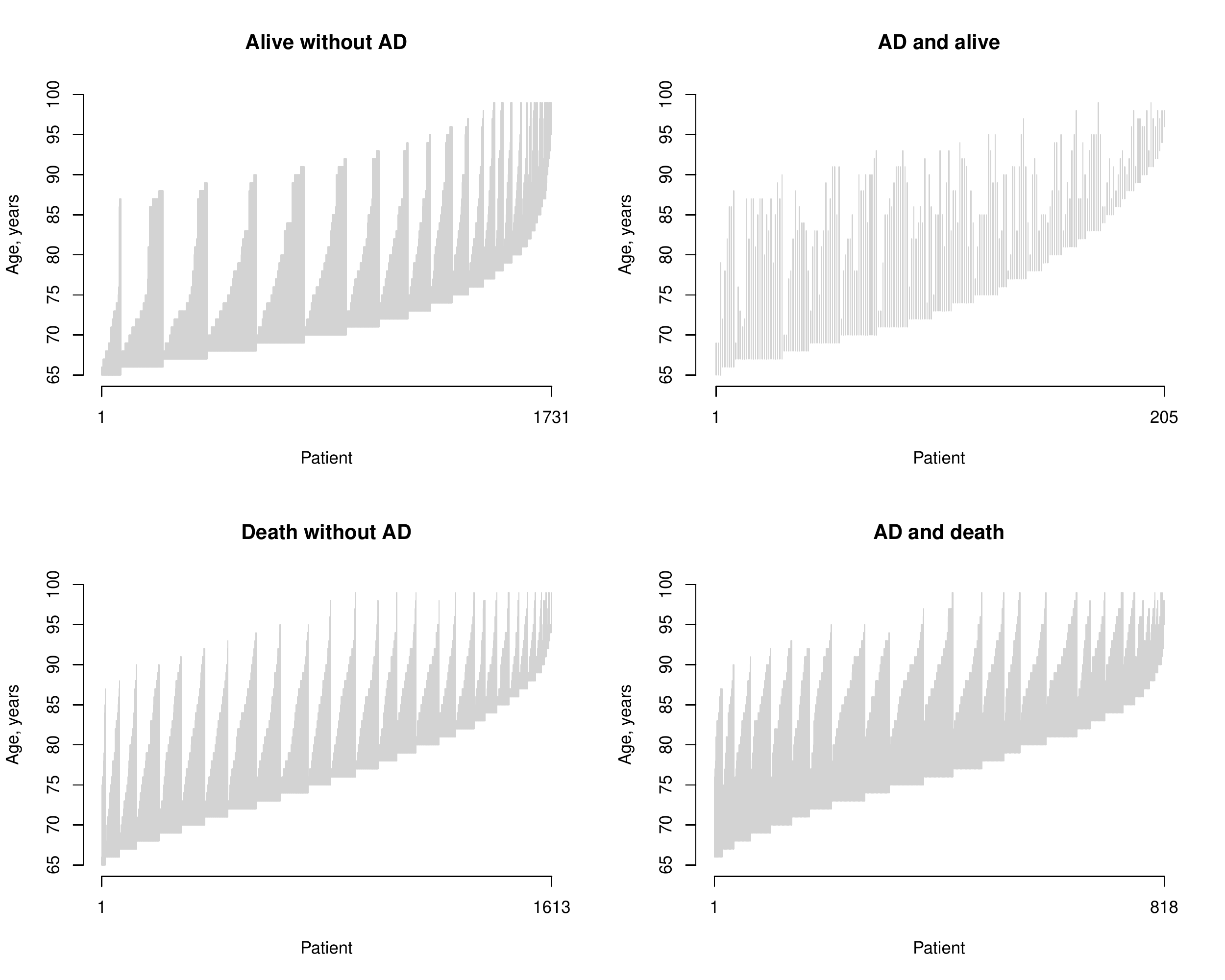}
\caption{\label{Fig:ACT:eda} Summary of person-time while on-study among $N$=4,367 participants in the ACT study, stratified by whether they had a diagnosis of AD and/or died during follow-up. Within each sub-figure, participants are ordered first by their age of enrollment and second by their observed event/censoring time.}
\end{center}
\end{figure}

Follow-up in ACT consists of biennial visits during which co-morbidities and clinical histories are updated, and during which participants undergo a comprehensive evaluation the result of which may be a diagnosis of AD. For the purposes of this paper, follow-up time was administratively censored at the first of December, 2016 or age 99 years. Based on these criteria 205 (5\%) were diagnosed with AD during follow-up but were censored prior to death, 818 (19\%) were diagnosed with AD and died during follow-up, 1,613 (37\%) died during follow-up without a diagnosis of AD; and, 1,731 (39\%) were censored prior to either a diagnosis of AD or experiencing death. Figure \ref{Fig:ACT:eda} provides a summary of the observed person-time for the patients. Within each panel, the patients have been ordered by: (i) their age at entry and (ii) the age at which their eventual outcome status is observed.

\section{A longitudinal bivariate modeling framework}
\label{Sec:ProposedFramework}

As indicated in the Introduction, a common theme across existing approaches to the analysis of semi-competing risks data is that they generally view dependence between the non-terminal and terminal events as a (statistical) nuisance and not a potential source of new clinical knowledge. In this Section we propose a novel longitudinal bivariate modeling framework for semi-competing risks within which dependence between $T_1$ and $T_2$ is characterized in a meaningful and interpretable way, and is permitted to be a function of covariates.
	
\subsection{A novel representation of semi-competing risks data}
\label{SubSec:NovelRep}

The proposed framework builds on a novel representation of observed semi-competing risks outcome data. Prior to describing this, it is useful to consider how such data are typically represented. Towards this, let $L$ and $C$ denote the study entry and right censoring times, respectively. The observed outcome data for the $i^{th}$ study participant is $\mathcal{D}_i = (L_i,\widetilde{T}_{1i},\delta_{1i},\widetilde{T}_{2i},\delta_{2i})$, where $\widetilde{T}_{1i} = \min(T_{1i},T_{2i},C_i)$, $\widetilde{T}_{2i} = \min(T_{2i},C_i)$, $\delta_{1i} = I\{\widetilde{T}_{1i}=T_{1i}\}$, and $\delta_{2i} = I\{\widetilde{T}_{2i}=T_{2i}\}$. Note, $\delta_{1i}$ and $\delta_{2i}$ indicate whether the study unit is observed to experience the non-terminal event and terminal event, respectively.

Let $\BFtau=\{\tau_0, \ldots, \tau_K\}$ be a set of user-specified points, with $\tau_{k-1} < \tau_k$ for $k = 1, \ldots, K$, that define a partition of the analysis time scale. Towards the study of AD using data from the ACT study, for example, one could choose age beyond 65 years as the time scale and $\BFtau=\{65,70,75,...,100\}$. Let $Y_{1i,k} = I\{T_{1i}\le \tau_k\}$ and $Y_{2i,k} = I\{T_{2i}\le \tau_k\}$ be indicators of whether participant $i$ experienced the non-terminal and terminal events by time $\tau_k$, respectively. Furthermore, let $k^{l}_i \in \{1, \ldots, K\}$ be the index of the first time point in $\BFtau$ after $L_i$ and $k^r_i \in \{1, \ldots, K\}$ the index of the first time point in $\BFtau$ after $\widetilde{T}_{2i}$. The observed outcome data on the partition $\BFtau$ can then be represented as a longitudinal bivariate process, specifically $\BFYi = \{(Y_{1i,k}, Y_{2i,k}); k = k^{l}_i, \ldots, k^r_i\}$. Figure A.1 in the Supplementary Materials provides a graphical representation of the observed outcome data for four hypothetical study participants.

Finally, let $\BFX_i(\tau_k)$ denote a vector of possibly time-dependent covariates measured at time $\tau_{k-1}$. Given this, let $\overline{\BFX}_{i,k} = (\BFX_i(\tau_1), \ldots, \BFX_i(\tau_k))$ denote the history of all covariate information (i.e. both time-invariant and time-varying) up to time point $\tau_{k-1}$. 

\subsection{The joint distribution of the observed data}
\label{SubSec:ProposedFramework:joint}

Let $\overline{\BFX}_i \equiv \overline{\BFX}_{i,{k^r_i}}$ is the totality of all covariate data observed for the $i^{th}$ participant. At the outset we assume that the joint distribution of the observed outcome data for the $i^{th}$ study participant, $\mP(\BFYi=\BFyi|\ \overline{\BFX}_i)$, can be decomposed as the product:
\begin{equation}
\label{Eq:MarkovAssump}
	\prod_{k=k^{l}_i}^{k^r_i}\mP(Y_{1i,k}=y_{1i,k},Y_{2i,k}=y_{2i,k}|\ Y_{1i,k-1}=y_{1i,k-1},Y_{2i,k-1}=y_{2i,k-1}, \overline{\BFX}_{i,k}),
\end{equation}
where we define $Y_{1i,0}=Y_{2i,0}=0$. Underpinning this decomposition is a Markov-type assumption that the joint probability of the two events in a given interval depends on the history of the two events solely through what is known at the start of the interval, conditional on the totality of the covariate information to that point.

As shown in Section A.4 of the Supplementary Materials, the components of expression (\ref{Eq:MarkovAssump}) can be written in terms of:
\begin{eqnarray}
\label{Eq:p1}
	\pi_{1i,k} &=& \mP(Y_{1i,k}=1|\ Y_{1i,k-1}=0, Y_{2i,k-1}=0, \overline{\BFX}_{i,k}) \\ 
\label{Eq:p2}
	\pi_{2i,k}(y_1) &=& \mP(Y_{2i,k}=1|\ Y_{1i,k-1}=y_1, Y_{2i,k-1}=0, \overline{\BFX}_{i,k}) \\ 
\label{Eq:theta}
	\theta_{i,k} &=& \mOR(Y_{1i,k},Y_{2i,k}|\ Y_{1i,k-1}=0, Y_{2i,k-1}=0, \overline{\BFX}_{i,k})
\end{eqnarray}
for $y_1=0,1$ and where
\[
	\mOR(Y_{1i,k},Y_{2i,k}|\ \ldots)\ =\ \frac{\mP(Y_{1i,k}=1, Y_{2i,k}=1|\ \ldots)\times\mP(Y_{1i,k}=0, Y_{2i,k}=0|\ \ldots)}{\mP(Y_{1i,k}=0, Y_{2i,k}=1|\ \ldots)\times\mP(Y_{1i,k}=1, Y_{2i,k}=0|\ \ldots)}.
\]
We emphasize that the interpretations of these three quantities are specific to the partition $\BFtau$. Specifically, $\pi_{1i,k}$ is the cumulative probability of experiencing the non-terminal event by time $\tau_k$, given that neither event had occurred by time $\tau_{k-1}$. Similarly, $\pi_{2i,k}(\cdot)$ is the cumulative probability of experiencing the terminal event by time $\tau_k$, given that the individual is alive at time $\tau_{k-1}$. Note, both of these correspond to quantities that are modeled in discrete time survival analyses \cite{lee2018analysis, prentice1978regression}. Finally, $\theta_{i,k}$ is the cross-sectional odds ratio for the 2$\times$2 table corresponding to the four possible observed ($Y_{1i,k}, Y_{2i,k}$) outcome vectors at time $\tau_k$, given that neither the non-terminal nor the terminal event had occurred by time $\tau_{k-1}$. Given its clear importance, we return to the choice of $\BFtau$ in Section \ref{Sec:ProposedFramework:choice}.

\subsection{Regression structure}
\label{SubSec:RegStruct}

We proceed with modeling by placing structure on the components given by expressions (\ref{Eq:p1})-(\ref{Eq:theta}) as a function of covariates. Specifically, we propose that $\pi_{1i,k}$, $\pi_{2i,k}(\cdot)$, and $\theta_{i,k}$ be modeled via the following regressions:
\begin{eqnarray}
\label{Eq:p1:model}
	\pi_{1i,k} &=& g_1^{-1}\{\alpha_{1,k}\ +\ f_1(\overline{\BFX}_{i1,k}; \BFbeta_1)\} \\ 
\label{Eq:p2:model}
	\pi_{2i,k}(y_1) &=& g_2^{-1}\{\alpha_{2,k}\ +\ f_2(\overline{\BFX}_{i2,k}, Y_{1i,k-1}; \BFbeta_2)\} \hspace*{0.3in} \\ 
\label{Eq:theta:model}
	\theta_{i,k} &=& g_\theta^{-1}\{\alpha_{\theta,k}\ +\ f_\theta(\overline{\BFX}_{i\theta,k}; \BFbeta_\theta)\},
\end{eqnarray}
where $g_1(\cdot)$, $g_2(\cdot)$ and $g_\theta(\cdot)$ are user-specified link functions (e.g. the logistic or log link), where $\overline{\BFX}_{i1,k}$, $\overline{\BFX}_{i2,k}$, $\overline{\BFX}_{i\theta,k}$ are each subsets of $\overline{\BFX}_{i,k}$, and where $f_1(\cdot)$, $f_2(\cdot)$ and $f_\theta(\cdot)$ are user-specified functions that characterize how the respective quantities depend on the covariates. For example, one may adopt a linear predictor specification with no interactions between the components of $\overline{\BFX}_{i2,k}$ and $Y_{1i,k-1}$ in the model for $\pi_{2i,k}(y_1) $ by setting $f_2(\overline{\BFX}_{i2,k}, Y_{1i,k-1}; \BFbeta_2) \equiv \overline{\BFX}_{i2,k}^T\BFbeta_{2,X} + Y_{1i,k-1}\beta_{2,y}$.

While their precise interpretations will depend on the partition of the time axis and the chosen link functions, $\BFbeta = (\BFbeta_1, \BFbeta_2, \BFbeta_\theta)$ characterize the impact of covariates; Section \ref{Sec:ProposedFramework:dependence} considers the role and interpretation of components of $\BFbeta_2$ that correspond to $Y_{1i,k-1}$. Finally, $\BFalpha_1$ = $(\alpha_{1,1}, \ldots, \alpha_{1,K})$, $\BFalpha_2$ = $(\alpha_{2,1}, \ldots, \alpha_{2,K})$, and $\BFalpha_\theta$ = $(\alpha_{\theta,1}, \ldots, \alpha_{\theta,K})$ represent baseline time trends in $\pi_{1i,k}$, $\pi_{2i,k}(\cdot)$, and $\theta_{i,k}$, respectively, with their precise interpretation again depending on the choice of $\BFtau$, the link functions and covariates included in the models.

From a practical perspective, that $\BFalpha$ = ($\BFalpha_1$, $\BFalpha_2$, $\BFalpha_\theta$) consists of 3$K$ parameters may result in computational and/or convergence issues unless the observed data is rich (i.e. a large sample size or high event rates) or the initial partition is coarse. To mitigate such issues we propose that a B-spline structure be adopted across the $K$ components of each of $\BFalpha_1$, $\BFalpha_2$ and $\BFalpha_\theta$~\citep{eilers1996flexible}. Towards the specification for $\BFalpha_1$, let $\{\tilde{t}_1^1, \ldots, \tilde{t}_1^{J_1}\}$ denote a collection of $J_1$ user-specified knots on the interior of the range ($\tau_0, \tau_K$) and $q_1$ the user-specified degree of the local polynomial basis functions. Given these choices, we specify the $k^{th}$ component of $\BFalpha_1$ as $\alpha_{1,k}=\sum_{j=1}^{\widetilde{J}_1}\eta_{1,j}B_{j,k},$ where $B_{j,k}$ is the value of the $j^{th}$ B-spline at time $\tau_k$ and $\widetilde{J}_1 = J_1-1+q_1$ is total number of spline terms. Thus, this specification requires estimation of $\widetilde{J}_1$ coefficient terms, specifically $\BFeta_1 = (\eta_{1,1}, \ldots, \eta_{1,\widetilde{J}_1})$. Applying the same strategy to $\BFalpha_2$ and $\BFalpha_\theta$ will result in $\widetilde{J}_1+\widetilde{J}_2+\widetilde{J}_\theta$ unknown coefficients, which we collectively denote as $\BFeta$ = ($\BFeta_1$, $\BFeta_2$, $\BFeta_\theta$). Finally, to distinguish between specifications, in the remainder of this paper we refer to the model based on $\BFalpha$ with 3$K$ unknowns as the \textit{unstructured model} while that based on $\BFeta$ with $\widetilde{J}_1+\widetilde{J}_2+\widetilde{J}_\theta$ unknowns as the \textit{B-spline model}. 

\subsection{Dependence}
\label{Sec:ProposedFramework:dependence}

The central innovation of the framework in expressions (\ref{Eq:p1:model})-(\ref{Eq:theta:model}) is that dependence between the non-terminal and terminal events is quantified in two distinct and yet complimentary ways. The first is captured through $\theta_{i,k}$ which can be viewed as a measure of \textit{local dependence} in that it quantifies the risk of co-occurrence of the two events, via the odds-ratio~\citep{prentice1978regression, lipsitz1991generalized, carey1993modelling, ten1999mixed, lee2018analysis, o2004analysis}, during the $(\tau_{k-1}, \tau_k]$ interval; a large positive value of $\theta_{i,k}$ indicates that if the non-terminal event (e.g. AD) occurs during $(\tau_{k-1}, \tau_k]$ then the terminal event (e.g. death) is likely to subsequently occur during the same interval. Because of the novel parameterization, we emphasize that the local dependence can vary over the time scale or as a function of covariates. As such, one could investigate whether local dependence between AD and death is weaker at younger ages relative to older ages. Furthermore, one could investigate whether the probability that the two events co-occur changes in response to key life events such as the death of a partner.

The second quantification of dependence is through the components of $\BFbeta_2$ in expression (\ref{Eq:p2:model}) that correspond to how the status of the non-terminal event, $Y_{1i,k-1}$, influences $\pi_{2i,k}$. Labelling these components as $\BFbeta_{2,y}$, analogous to how one interprets covariates effects in regression models, these parameters capture the extent to which whether the non-terminal event has occurred is associated with a change in risk of the terminal event; for this reason, we refer to $\BFbeta_{2,y}$ as capturing \textit{global dependence}. Note, global dependence is conceptually similar to the explanatory hazard ratio (EHR) and cross-quantile residual ratio (CQRR)~\cite{yang2016new} in that it concerns the role that the non-terminal event plays in modifying the future occurrence of the terminal event (see Section A.1 of the Supplementary Materials).

\subsection{Choice of partition}
\label{Sec:ProposedFramework:choice}

The partition, $\BFtau$, plays a critical role in the proposed framework in that it provides the foundation for being able to distinguish between local and global dependence, as we have conceptualized them, and for being able to investigate the role that covariates play. In Section A.4 of the Supplementary Materials we show that a distribution for $(T_1, T_2)$ induces the quantities \eqref{Eq:p1:model}--\eqref{Eq:theta:model} for \textit{any} $\BFtau$. Thus, regardless of the choice of $\BFtau$, the components of the proposed model are well-defined mathematical objects and are, therefore, valid targets for estimation and inference. An important consequence of this is that one cannot say that any given partition corresponds to the `truth'. As indicated in Section \ref{SubSec:ProposedFramework:joint}, however, the choice of $\BFtau$ dictates the numerical values and interpretation of the quantities given by (\ref{Eq:p1})-(\ref{Eq:theta}), and correspondingly (in part, at least) the numerical values and interpretation of the parameters in the regression structure given by expressions (\ref{Eq:p1:model})-(\ref{Eq:theta:model}). As such the choice of $\BFtau$ is a critical challenge that requires careful consideration by the study investigators.

In principle, one could approach choosing $\BFtau$ through consideration of the clinical condition under investigation such as the pace at which the disease progresses. In the ACT study, for example, the choice to schedule follow-up biennially was made in part for logistical considerations (it would be challenging to follow-up approximately 4,000 participants each year) but also because AD is a slowly-developing condition. Alternatively, one may pursue a data-driven strategy where, for example, some goodness-of-fit criterion is specified and then optimized as a function of $\BFtau$. Our perspective is that the decision should be based primarily, if not exclusively, on clinical considerations. Central to this position is that, in addition to their interpretation, the numerical values of (\ref{Eq:p1})-(\ref{Eq:theta}) change with $\BFtau$. To see this, we again note that the probabilities in each of (\ref{Eq:p1})-(\ref{Eq:theta}) speak to the cumulative incidence of events during the interval $(\tau_{k-1}, \tau_k]$. If the length of the interval is decreased then the incidence will necessarily decrease. The corresponding new interpretation and numerical value will not be `wrong', however, but just different. Put another way, the change in the interpretation and the numerical values of the model parameters that result from, say, adopting a finer partition of the time scale should, arguably, be viewed as a change in the question that is being answered. Thus, in our view, purely data-driven approaches, while they may have some initial intuitive appeal, should be avoided.

Nevertheless, we do acknowledge that it may not be easy to elicit a single partition, from the literature or collaborators, on which to base the analyses and conclusions. If it is the case that there is no clear choice, analysts may opt to perform a range of analyses over different partitions, both in terms of how fine the partition is and in terms of where the cut-points are for a given (common) interval length. We pursue this strategy in conducting the analyses of the data from ACT in Section \ref{Sec:ACTanalysis}.

\section{Estimation and inference}
\label{Sec:EstimationInference}
	
\subsection{The observed data likelihood}
\label{SubSec:ObsLik}

Building on the notation developed in Section \ref{Sec:ProposedFramework}, the first two columns of Table \ref{Tab:DataPatterns} provide the six possible outcome data scenarios in the $k^{th}$ interval of the partition given by $\BFtau$ as a function of the outcome vector in the previous interval (i.e. as a function of ($Y_{1i,k-1}, Y_{2i,k-1}$)). The third column provides the corresponding likelihood contributions, that is the interval-specific components in the decomposition given by expression (\ref{Eq:MarkovAssump}), with
\begin{eqnarray*}
	\pi_{12i,k} &=& \mP(Y_{1i,k}=1,Y_{2i,k}=1| Y_{1i,k-1}=0, Y_{2i,k-1}=0, \overline{\BFX}_{i,k}) \\
\label{Eq:P12calc}
	&=& \left\{\begin{array}{cc}
		\pi_{1i,k}\pi_{2i,k}(0) & \hspace*{0.2in} \theta_{i,k}=1 \\
		\frac{1}{{2(\theta_{i,k}-1)}}\times \left[1 + a_{ij}-\sqrt{(1 + a_{i,k})^2-4\theta_{i,k}(\theta_{i,k}-1)	\pi_{1i,k}\pi_{2i,k}(0)}\right] & \hspace*{0.2in} \theta_{i,k}\ne 1
\end{array}\right. ,
\end{eqnarray*}
where $a_{i,k}= (\pi_{1i,k} + \pi_{2i,k}(0))(\theta_{i,k}-1)$~\citep{ten1999mixed}. 

\begin{table}[t!]
\begin{center}
\centering
\begin{tabular}{ccc}
\hline\hline
	$(Y_{1i,k-1}, Y_{2i,k-1})$	& $(Y_{1i,k}, Y_{2i,k})$		& Likelihood contribution\\
\hline
	(0, 0)		& (0, 0)	& $1\ -\ \pi_{1i,k}\ -\ \pi_{2i,k}(0)\ +\ \pi_{12i,k}$ \\
	(0, 0)		& (1, 0)	& $\pi_{1i,k} - \pi_{12i,k}$ \\
	(0, 0)		& (0, 1)	& $\pi_{2i,k}(0) - \pi_{12i,k}$ \\
	(0, 0)		& (1, 1)	& $\pi_{12i,k}$ \\
	(1, 0)		& (1, 0)	& $1 - \pi_{2i,k}(1)$ \\
	(1, 0)		& (1, 1)	& $\pi_{2i,k}(1)$ \\
\hline\hline
\end{tabular}
	\caption{\label{Tab:DataPatterns} \footnotesize Six possible data scenarios for the $k^{th}$ interval in the partition given by $\BFtau$ (see Section \ref{Sec:ProposedFramework})}
\end{center}
\end{table} 

Let $\bphi$ denote the collection of unknown parameters in the specification of model (\ref{Eq:p1:model})-(\ref{Eq:theta:model}). Note, if the unstructured form of the model is fit then $\bphi\equiv\bphi^{\balpha}=(\balpha, \bbeta)$ while $\bphi\equiv\bphi^{\BFeta}=(\BFeta, \bbeta)$ if the B-spline model is fit. For either specification, the observed data likelihood for a random sample of $N$ study participants from the population of interest, $\mathcal{L}(\bphi)$ is the product of $N$ terms, each of the form:
\begin{align*}
	\mathcal{L}_i(\bphi)\ & =\ \prod_{k=k^{l}_i}^{k^r_i}\mP(Y_{1i,k}=y_{1i,k},Y_{2i,k}=y_{2i,k}|\ Y_{1i,k-1}=y_{1i,k-1},Y_{2i,k-1}=y_{2i,k-1}, \overline{\BFX}_{i,k}), \\
	& =\ \prod_{k=k^{l}_i}^{k^r_i}\Big\{[\pi_{12i,k}]^{y_{1i,k}y_{2i,k}}[\pi_{1i,k}-\pi_{12i,k}]^{y_{1i,k}(1-y_{2i,k})} 
[\pi_{2i,k}(0)-\pi_{12i,k}]^{(1-y_{1i,k})y_{2i,k}} \\
	& \hspace*{0.5in} \times\
	 [1-\pi_{1i,k}-\pi_{2i,k}(0)+\pi_{12i,k}]^{(1-y_{1i,k})(1-y_{2i,k})}\Big\}^{(1-y_{1i,k-1})(1-y_{2i,k-1})} \\
	& \hspace*{0.5in} \times \Big\{[\pi_{2i,k}(1)]^{y_{2i,k}}[1-\pi_{2i,k}(1)]^{1-y_{2i,k}}\Big\}^{y_{1i,k-1}(1-y_{2i,k-1})}.
\end{align*}

\subsection{Estimation}
\label{SubSec:Estimaton}

If the unstructured form of model (\ref{Eq:p1:model})-(\ref{Eq:theta:model}) is adopted, estimation can proceed in a straightforward manner by finding the maximizer of $\ell(\bphi^{\balpha})$ = $\log \mathcal{L}(\bphi^{\balpha})$, denoted by $\widehat{\bphi}^{\balpha}$. Under the proposed B-spline model, however, care is needed to avoid overfitting. To resolve this, we maximize the observed data likelihood subject to a penalty that imposes smoothness in resulting estimated functions. One such penalty is the integrated squared second derivative which, following Eilers et al.~\cite{eilers1996flexible}, can be approximated by penalizing coefficient differences. Let $\Delta$ be the difference operator so that, for example, $\Delta \eta_{1,j}=\eta_{1,j}-\eta_{1,j-1}$. Then, letting $\blambda = (\lambda_1, \lambda_2, \lambda_\theta)$, consider the penalized likelihood:
\[
	\mathcal{L}(\bphi^{\BFeta}; \blambda)\ =\ \mathcal{L}(\bphi^{\BFeta})\ -\ \lambda_1\sum_{j=m+1}^{\widetilde{J}_1}(\Delta^m\eta_{1,j})^2\ -\ \lambda_2\sum_{j=m+1}^{\widetilde{J}_2}(\Delta^m\eta_{2,j})^2\ -\ \lambda_\theta\sum_{j=m+1}^{\widetilde{J}_\theta}(\Delta^m\eta_{\theta,j})^2,
\]
where $\Delta^m$ corresponds to the difference operator having been applied $m$ times; for example, $\Delta^2 \eta_{1,j}=\Delta (\Delta \eta_{1,j}) = \eta_{1,j}-2\eta_{1,j-1}+\eta_{1,j-2}$. Note, letting $\bD_m$ denote the matrix representation of $\Delta^m$, such that, for example, $\sum_{j=m+1}^{\widetilde{J}_1}(\Delta^m\eta_{1,j})^2$ can be written as $\BFeta_1^T \bD_m^T\bD_m \BFeta_1$ (Section A.4 of the Supplementary Materials), the penalized likelihood can be written as:
\begin{equation}
\label{Eq:PenalLikMatt}
	\mathcal{L}(\bphi^{\BFeta}; \blambda)\ =\ \mathcal{L}(\bphi^{\BFeta})\ -\ \lambda_1\BFeta_1^T \bD_m^T\bD_m \BFeta_1\ -\ \lambda_2\BFeta_2^T \bD_m^T\bD_m \BFeta_2\ -\ \lambda_\theta\BFeta_\theta^T \bD_m^T\bD_m \BFeta_\theta.
\end{equation}

Let $\ell(\bphi^{\BFeta}; \blambda) = \log\mathcal{L}(\bphi^{\BFeta}; \blambda)$ be the log-penalized likelihood. Furthermore, let $\bU(\bphi^{\BFeta};\blambda) = \nabla_{\bphi^{\BFeta}}\ell(\bphi^{\BFeta}; \blambda)$ be the gradient of $\ell(\bphi^{\BFeta}; \blambda)$ with respect to $\bphi^{\BFeta}$ and $\mathcal{H}(\bphi^{\BFeta};\blambda)=\nabla_{\bphi\bphi^{\BFeta}}\ell(\bphi^{\BFeta}; \blambda)$ the corresponding matrix of second partial derivatives (i.e. the Hessian matrix). For a given value of $\blambda$, the penalized maximum likelihood estimator, which we denote by $\widehat{\bphi}^{\BFeta}_{\blambda}$, is the solution to $\bU(\bphi^{\BFeta};\blambda)=\bzero$. Note, since the penalty is quadratic in $\BFeta_1$, $\BFeta_2$, and $\BFeta_\theta$, gradient-based maximization is relatively straightforward to carry out.

Finally, towards choosing the value of $\blambda$ on which the final results are to be based, say $\blambda^*$, one can proceed using any standard model-selection criteria such as the Akaike Information Criterion (AIC) or cross validation \citep{gray1992flexible, eilers1996flexible}. In our implementation of the methods, we follow Gray~\cite{gray1994spline} by taking the trace of $\mathcal{H}(\bphi^{\BFeta}; \bzero)\mathcal{H}^{-1}(\bphi^{\BFeta}; \blambda)$ as the effective degrees of freedom when calculating the AIC.
	
\subsection{Asymptotic properties}
\label{SubSec:AsyTheory}

For the unstructured model, under standard regularity conditions, the asymptotic distribution for the maximum likelihood estimator  $\widehat{\bphi}^{\balpha}$ follows from standard likelihood theory. That is, $\widehat{\bphi}^{\balpha}$ is consistent and, letting $\widetilde{\bphi}^{\balpha}$ denote the value of ${\bphi}^{\balpha}$ induced from the partition,   $N^{1/2}(\widehat{\bphi}^{\balpha}-\widetilde{\bphi}^{\balpha})$ is asymptotically normal with the variance being the usual 
inverse of Fisher information matrix. 

For the B-spline model, careful consideration of the asymptotic properties of $\widehat{\bphi}^{\BFeta}$ requires specification of how the number of knots, $\widetilde{\bJ} = (\widetilde{J}_1, \widetilde{J}_2, \widetilde{J}_\theta)$, and the penalty, $\blambda^*$, change with the sample size $N$. This has been the focus of a large body of work, much of which is concerned with error estimation for the non-parametric function(s)~\citep[e.g.,][]{li2008asymptotics, kauermann2009some, claeskens2009asymptotic}. Here, however, primary interest lies with statistical inference for the parameters $\bphi^{\BFeta}$. Given this, and as recommended by number of authors for development of practical inference in a range of setting~\cite{gray1992flexible, gray1994spline, wang1995inference, yu2002penalized}, we consider the behavior of $\widehat{\bphi}^{\BFeta}$ in settings where both $\widetilde{\bJ}$ and $\blambda$ are fixed. With this, henceforth, we denote the estimator as $\widehat{\bphi}^{\BFeta}_{\blambda}$.

To this end, we show in  Section A.5 of the Supplementary Materials, that under standard regularity assumptions $N^{1/2}(\widehat{\bphi}^{\BFeta}_{\blambda}-\widetilde{\bphi}^{\BFeta}_{\blambda})$ is asymptotically normal with variance that can be consistently estimated by 
\begin{equation}
\label{Eq:VarEst}
	\widehat{\mathcal{V}}\ =\ N\mathcal{H}^{-1}(\widehat{\bphi}^{\BFeta}; \blambda)\left(\frac{1}{N}\sum_{i=1}^{N}\bU_i(\widehat{\bphi}^{\BFeta};\blambda)\bU_i^T(\widehat{\bphi}^{\BFeta};\blambda)\right)N\mathcal{H}^{-1}(\widehat{\bphi}^{\BFeta};\blambda),
\end{equation}
where $\widetilde{\bphi}^{\BFeta}_{\blambda}$ is the solution of $E[\bU(\bphi^{\BFeta}; \blambda)]=\bzero$ and where $\bU_i(\cdot;\blambda)$ is the gradient of the log penalized likelihood of a single observation $i$. For individual parameters, such as the coefficient of APOE in the model for $\theta$, the appropriate entry in the diagonal of \eqref{Eq:VarEst} is used for estimating the variance. For the baseline time trends, let $\widehat{\mathcal{V}}_{\BFeta_1}$ be the sub-matrix corresponding to the estimated variance matrix of $\widehat{\BFeta}_1$. Note that $\widehat{\balpha}_1$ can be written as $\widehat{\balpha}_1=\widehat{\BFeta}^T_1\mathcal{B}$ with $\mathcal{B}$ being a $\widetilde{J}_1\times K$ matrix with elements $\mathcal{B}_{jk}$=$B_{j,k}$. Therefore, to construct a 95\% confidence interval for the AD time-varying reference probability, $\pi_{1i,k}$, one can use $g^{-1}_1\{\widehat{\BFeta}^1\pm1.96[diag(\mathcal{B}^T\widehat{\mathcal{V}}_{\BFeta_1}\mathcal{B})]^{1/2}\}$ where here $g^{-1}$ to be understood as operating entrywise on the vector, and $diag(\cdot)$ returns the diagonal of a matrix. 

\section{Simulation studies}
\label{Sec:Sims}

We conducted a series of simulations to investigate: (i) finite sample properties of the methods proposed in Section \ref{Sec:EstimationInference}; and, (ii) the potential bias-variance trade-off that analysts will have to contend with when choosing the degree of regularization in $\mathcal{L}(\bphi; \blambda)$. We note that the simulations make no attempt to perform a comparison with existing methods (e.g. in terms of bias or efficiency) since the proposed framework was developed to investigate dependence in a way that is distinct, and thus complimentary, from how existing methods approach it. We do, however, explore different approaches in the analysis of the ACT data in Section \ref{Sec:ACTanalysis}. Due to space constraints, we present brief details and a summary of the main conclusions; see Section A.6 in the Supplementary Materials for full details and results. 

Building on features of the observed data from the ACT study, we generated the data according to models \eqref{Eq:p1:model}--\eqref{Eq:theta:model}, as a function of both time-fixed and time-dependent covariates, under three scenarios for dependence: a null scenario, a simple dependence scenario and a complex dependence scenario.  A range of right-censoring rates (0-30\%) and sizes (500-5,000) were considered. For each scenario 1,000 datasets were generated. All simulations were carried out using code available in the \texttt{LongitSemiComp} package for \texttt{R}. Simulation code, seeds and results are all available online via the Github repository of the first author.

Let $\expit(v)=\exp(v)/[1+\exp(v)]$. Figures A.2 and A.3 in the Supplementary Materials summarize performance regarding estimation of time-varying functions. The time-varying terminal event probability function $\expit(\BFalpha_2)$ was well-estimated, regardless of the choice for number of knots and the penalty level. The time-varying non-terminal probability function $\expit(\BFalpha_1)$ was also well-estimated except when the number of knots was small and a substantial amount of regularization were used ($\widetilde{J}=5$ and $\lambda \ge5$) where the B-spline estimator was oversmoothed, resulting in bias for later time points where less information is available. The time-varying odds ratio function $\exp(\BFalpha_\theta)$ was well estimated by the 10 knots B-spline estimator, as well as by the B-spline non-smoothed 5 knots B-spline estimator. Similar to the non-terminal probability, the 5 knots oversmoothed estimator suffered from bias. A model with completely unrestricted $\BFalpha_1$, $\BFalpha_2$, and $\BFalpha_\theta$ worked well until the last time point, where some bias was observed. Instability of the undersmoothed estimators for time-varying odds ratio, and biased, yet stable, estimators when using excessive smoothing were found when the sample size was small. Using AIC to choose $\lambda$, more smoothing was desired for $\widetilde{J}=10$ compared with $\widetilde{J}=5$ (Table A.4). 

Turning to the coefficients (Tables A.5-A.12), small finite-sample bias for $\bbeta_1$ and $\bbeta_2$ was mitigated under larger sample sizes. The global dependence parameter $\BFbeta_{2,y}$ was well-estimated, with negligible bias for all sample sizes and censoring rates. A small finite-sample bias was observed for $\bbeta_\theta$ in some of the scenarios, although it decreased as the sample size increased. Furthermore, the bias was more substantial when the sample size was low and the unrestricted model was used for the time-varying functions. Unpenalized estimation under B-spline representation also resulted in bias for $\bbeta_\theta$. However, penalization largely mitigated this bias and reduced the standard error. For larger sample sizes, this bias disappeared. These results were consistent with the performance of the time-varying component estimator of the odds ratio. That is, when $\BFalpha_\theta$ was estimated well, so was $\bbeta_\theta$. Finally, in most cases our variance estimators performed very well, and the empirical coverage of the confidence intervals was close to the desired nominal level.

\section{Analysis of data from the ACT study}
\label{Sec:ACTanalysis}

Having $\ge$ 1 APOE $\epsilon$4 allele is well-established as a genetic risk factor for AD~\cite{baumgart2015summary}. The extent to which having $\ge$ 1 APOE $\epsilon$4 allele is associated with mortality, however, is unclear~\cite{helzner2008survival}; results across published studies conflict, with some indicating shorter survival but only among men~\cite{dal2002apoe}, some indicating no association~\cite{koivisto2000apolipoprotein} and, others indicating prolonged survival~\cite{van1995apolipoprotein}. To the best of our knowledge, however, none of the previous studies have examined the role of the APOE $\epsilon$4 allele through the lens of semi-competing risks. As such, the opportunities that semi-competing risks analyses provide, particularly in terms of explicit acknowledgement of death as a competing risk and in terms of being able to learn about dependence, have not been taken advantage of. 

With this backdrop, we present a detailed case study with the goal to investigate the role of having $\ge$ 1 APOE $\epsilon$4 allele on the joint risk of AD and death, and whether this varies by gender. Throughout, we use data from ACT with the time scale taken to be `time since age 65' or `time since a diagnosis of AD', as appropriate. In reporting results we focus on those that pertain to dependence; additional details, results and sensitivity analyses are provided in the Supplementary Materials.

\subsection{Analyses based on existing methods}	 
\label{Sec:ACTanalysis:existing}

Prior to presenting results based on the proposed framework, we consider a series of analyses based on existing methods described in  Section A.1 of the Supplementary Materials. Results are given in Figure \ref{Fig:Existing}, and in Section A.6 of the Supplementary Materials.

\subsubsection{An illness-death model with a patient-specific frailty}

At the outset we considered the illness-death framework, with the hazard functions  for the three transitions (i.e. Healthy $\Rightarrow$ AD, Healthy $\Rightarrow$ Death and AD $\Rightarrow$ Death) specified via the following Cox-type models  (Section A.1):
\begin{eqnarray}
\label{Eq:ID1}
	\lambda_1(t_1; \BFX_i) &=& \gamma_i\lambda_{01}(t_1)\exp\{\BFX_i^T\BFxi_1\}, \\
\label{Eq:ID2}
	\lambda_2(t_2; \BFX_i) &=& \gamma_i\lambda_{02}(t_2)\exp\{\BFX_i^T\BFxi_2\}, \\
\label{Eq:ID3}
	\lambda_3(t_2| t_1; \BFX_i) &=& \gamma_i\lambda_{03}(t_2 - t_1)\exp\{\BFX_i^T\BFxi_3\},
\end{eqnarray}
where $\gamma_i$ $\sim$ Gamma($\theta^{-1}$, $\theta^{-1}$) is a patient-specific frailty~\cite{lee2015bayesian, xu2010statistical}. Note, the model for $\lambda_3(t_2| t_1; \cdot)$ is \textit{semi-Markov}; the time scale is time since diagnosis of AD. For all three transitions, the baseline hazard is modeled as a Weibull($\nu_g, \kappa_g$), such that $\lambda_{0g}(t)$ = $\nu_g\kappa_gt^{\nu_g-1}$. These choices were made because, as far as we are aware, there are no implementations of the illness-death model, as specified by expressions (\ref{Eq:ID1})-(\ref{Eq:ID3}), that permit left truncation (a key feature of the ACT data) and anything other than Weibull baseline hazards.

Tables A.13 and A.14, and Figure A.4 in the Supplementary Materials report results for four illness-death models, that differ in whether a patient-specific frailty was incorporated and whether age at AD diagnosis was included in $\lambda_3(t_2| t_1; \cdot)$. A key observation from these analyses is that there is little evidence that the frailties, as they are included in models (\ref{Eq:ID1})-(\ref{Eq:ID3}), serve to account for any of the dependence between the two events above and beyond how dependence is structured through the interplay of the remaining components of the illness-death model: the point estimates for log($\theta$) are $-$14.34 and $-$13.10, and the log-likelihood at the maximum likelihood estimates are the same whether one includes the frailties or not (Table A.14). Moreover, that the point estimates for $\theta$ are (essentially) on the boundary of the parameter space results in the hessian evaluated at the maximum likelihood estimates not being invertible.

\subsubsection{An illness-death model with no patient-specific frailty}

Motivated by the above results, we performed additional analyses based on Cox-type specifications for the three transition-specific hazards but without the $\gamma_i$ frailties. Note, in removing the frailties one can estimate model components using partial likelihood methods for Cox models where the data are subject to left truncation. We fit two semi-Markov models, as in expressions (\ref{Eq:ID1})-(\ref{Eq:ID3}), again differing in whether age at AD diagnosis was included. We also fit a \textit{Markov} model in which the AD $\Rightarrow$ Death transition is modeled as:
\[
	\lambda_3(t_2| t_1; \BFX_i)\ =\ \lambda_{03}(t_2)\exp\{\BFX_i^T\BFxi_3\},
\]
so that the time scale is the same as for the other two transitions (i.e. time since age 65).

Tables A.15 and A.16 in the Supplementary Materials report estimated hazard ratios and 95\% confidence intervals. Overall, the conclusions regarding the associations between the covariates and both Alzheimer's disease and mortality are consistent with the Weibull illness-death model fits. Also consistent is the evidence regarding the dependence between AD and mortality as quantified by the inclusion of age at AD diagnosis in the model for $\lambda_3(t_2| t_1; \BFX_i)$; the estimated hazard ratio for a 5-year contrast is 1.33 (95\% CI: 1.24, 1.42).

From the Markov model, since Healthy $\Rightarrow$ Death and the AD $\Rightarrow$ Death transitions are modeled on the same time scale, one can report the explanatory hazard ratio for any given patient profile. The top-left sub-figure of Figure \ref{Fig:Existing} provides smoothed estimates of EHR($t_2$; $t_1$) for four profiles (Figure A.5 in the Supplementary Materials provides additional detail). From the figure it is clear that, at any given age, the hazard for death is substantially higher for individuals who have been diagnosed with AD relative to those who have not, with the biggest differences among the relatively young. Intuitively, this suggests that, among individuals at least 65 years of age, a diagnosis of AD is more devastating (from a mortality perspective) for a young individual than for an older individual.

\begin{figure*}[t!]
\centering
\begin{subfigure}[t]{0.5\textwidth}
	\centering
	\includegraphics[width=3in]{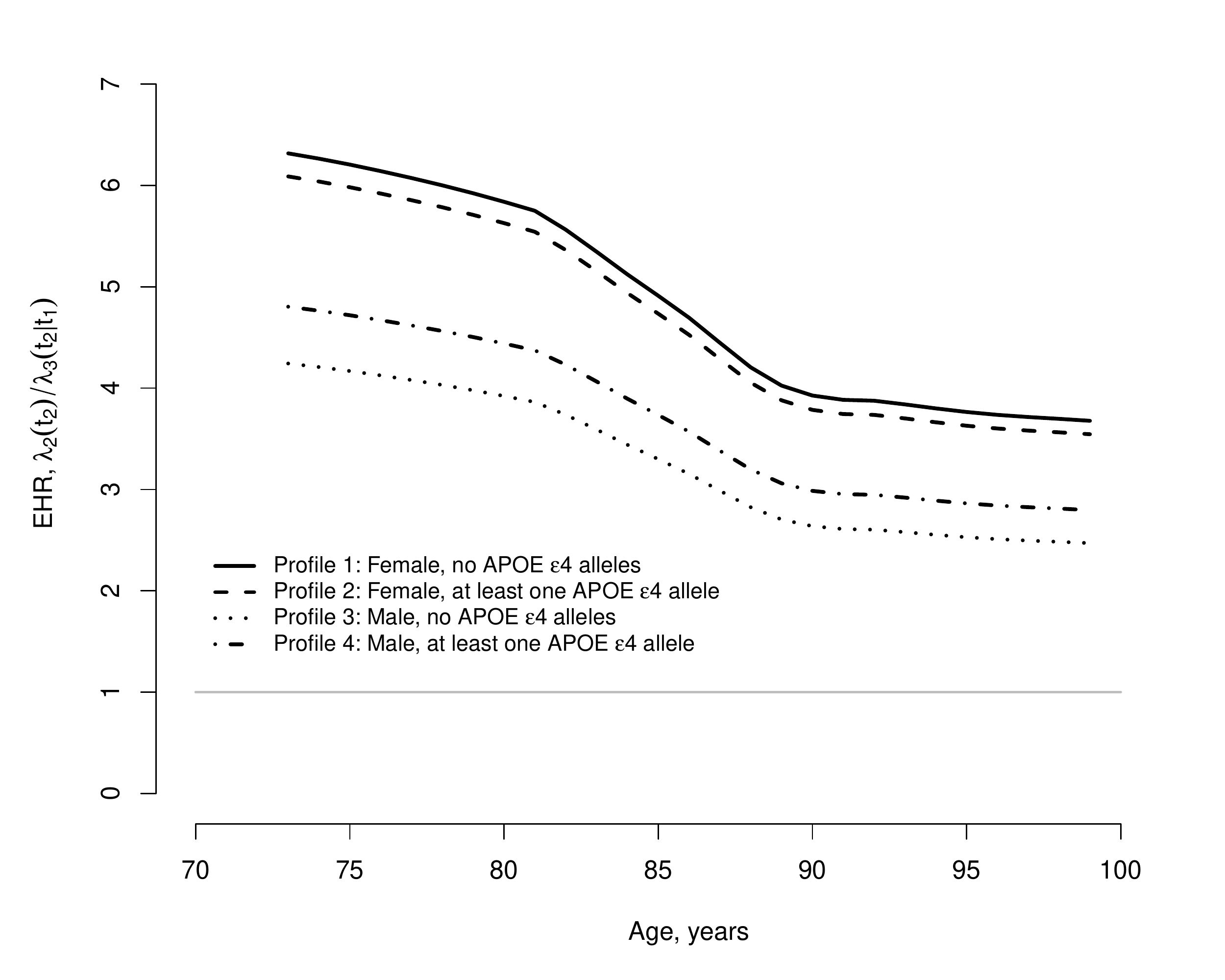}
\end{subfigure}%
\begin{subfigure}[t]{0.5\textwidth}
	\centering
	\includegraphics[width=3in]{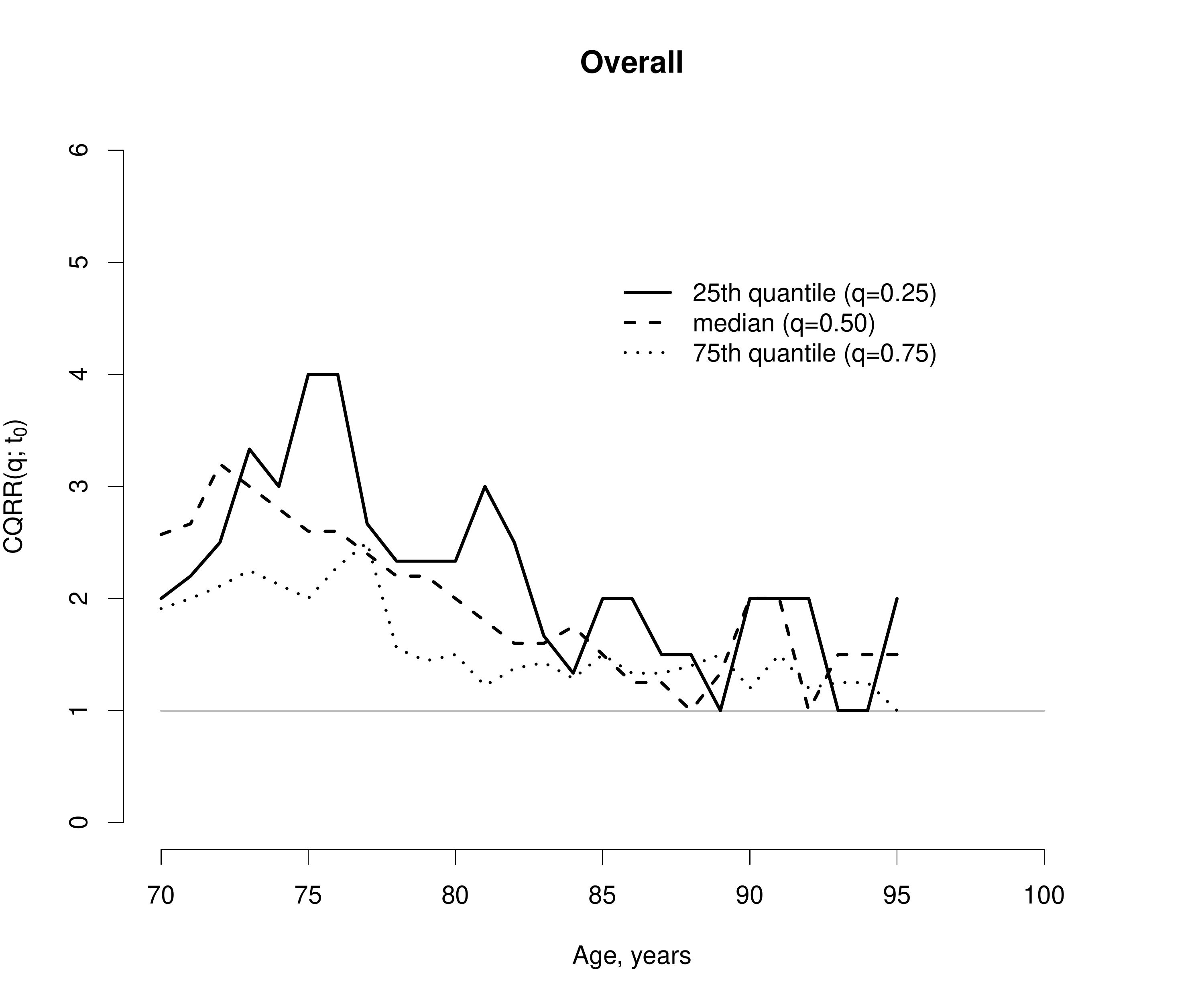}
\end{subfigure}
~~
\begin{subfigure}[t]{0.5\textwidth}
	\centering
	\includegraphics[width=3in]{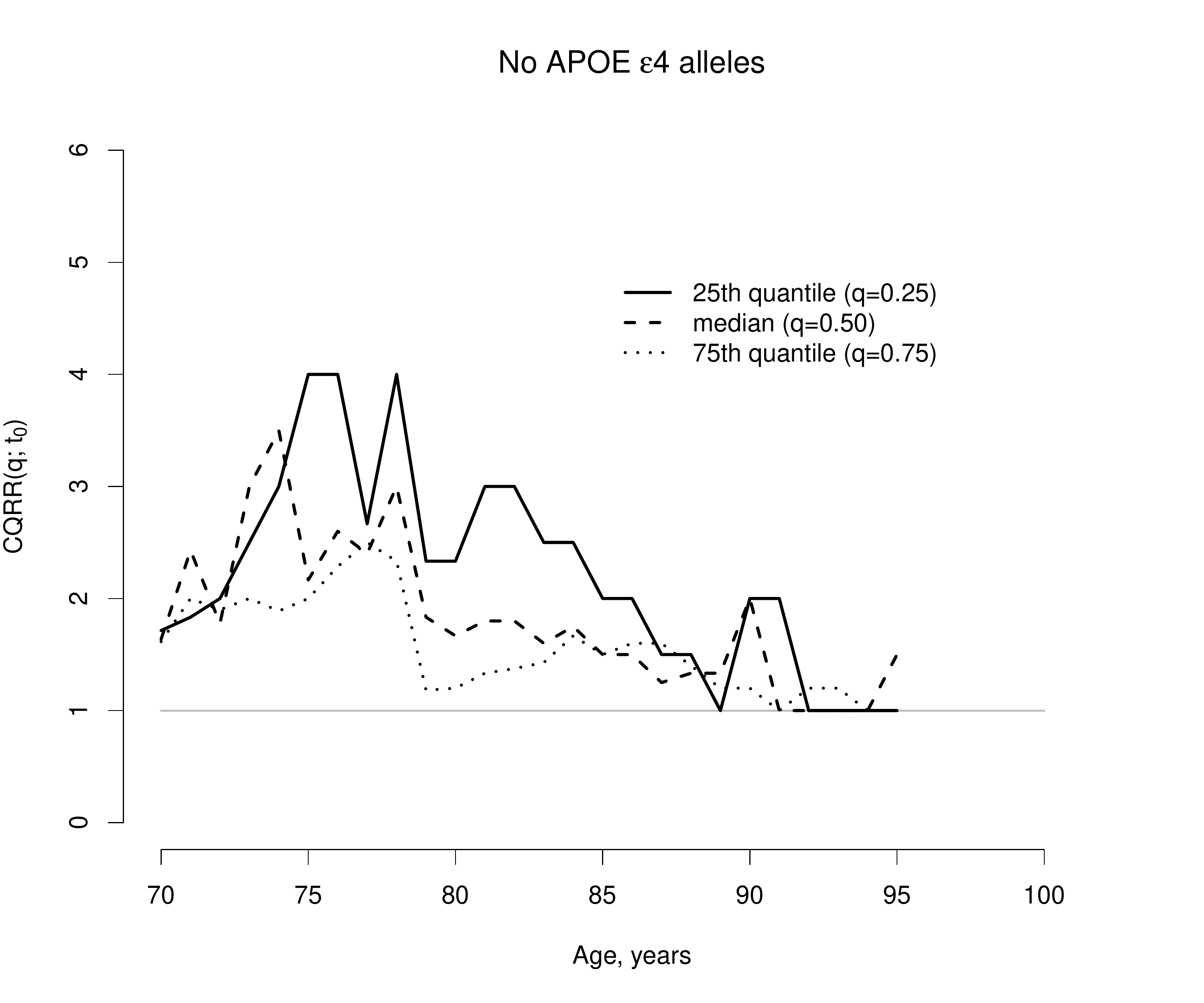}
\end{subfigure}%
\begin{subfigure}[t]{0.5\textwidth}
	\centering
	\includegraphics[width=3in]{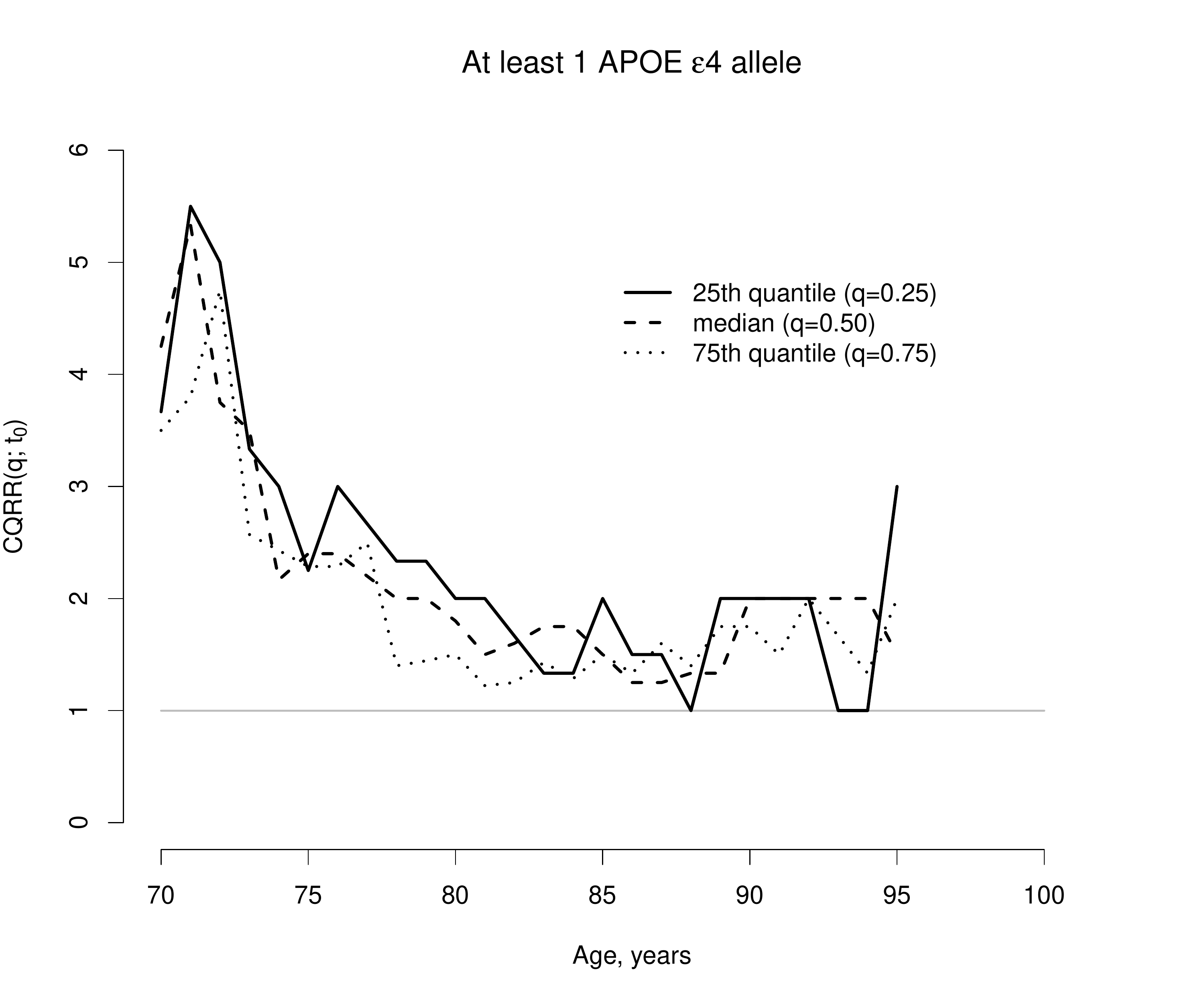}
\end{subfigure}
\caption{Estimated EHRs for four patient profiles and CQRRs for the 25th quantile, the median and 75th quantile of residual time for death, overall and stratified by whether there was at least 1 APOE $\epsilon$4 allele, based on analyses of the ACT data. See Section \ref{Sec:ACTanalysis:existing}. \label{Fig:Existing}}
\end{figure*}

\subsubsection{The cross-quantile residual ratio}

Finally, we report results based on the CQRR methodology\footnote{Using code available at http://web1.sph.emory.edu/users/lpeng/Rpackage.html}. Specifically, we calculated the CQRR($q; t_0$)  at $q \in \{0.25,0.5.0.75\}$ (i.e. for the 25th quantile, the median and 75th quantile of residual time for death) for ages $t_0 \in \{70, \ldots, 95\}$ years, stratifying by the number of APOE $\epsilon$4 alleles (0 vs $\ge$ 1).

Figure \ref{Fig:Existing} provides estimates; see Figure A.6 in the Supplementary Materials for 95\% confidence intervals. From Figure \ref{Fig:Existing}, the estimated CQRR($\tau; t_0$) are greater than 1.0 at all ages, indicating that residual lifetime, at any given age, for individuals without a diagnosis of AD is estimated to be longer than that for individuals with an AD diagnosis. Comparing the bottom sub-figures of Figure \ref{Fig:Existing}, we see that the spread in the lines is somewhat greater for patients with no APOE $\epsilon$4 alleles. To interpret this, consider the population of patients with no APOE $\epsilon$4 alleles and the population with at least one. In comparing patients without an AD diagnosis to those with such a diagnosis, the distribution of residual lifetime exhibits less variation, at any given age, in the second of these populations. Thus, a diagnosis of AD in patients with at least one APOE $\epsilon$4 allele results in a relatively homogeneous decline whereas the decline associated with a diagnosis of AD in patients with no APOE $\epsilon$4 alleles is more heterogeneous.

\subsection{Analyses based on the proposed framework}
\label{Sec:ACTanalysis:proposed}

Finally, we report results from a series of analyses using the proposed longitudinal bivariate modeling framework. As emphasized in Section \ref{Sec:ProposedFramework:choice}, the choice of the partition $\BFtau$ is important. Towards investigating the role of the choice, and recalling from Section \ref{Sec:ACT} that participants underwent biennial visits, we considered two partitions of the time scale $[65, 100)$: $\BFtau^{2.5} = \{65.0, 67.5, 70.0, \ldots, 97.5, 100.0\}$, for which $K$=14, and $\BFtau^{5.0} = \{65.0, 70.0, 75.0 \ldots, 95.0, 100.0\}$, for which $K$=7. Tables A.2 and A.3 in the Supplementary Materials provide the $2\times2$ outcome tables for each interval, under the two participations.

Towards specification of the models in expressions (\ref{Eq:p1:model})-(\ref{Eq:theta:model}), we used logit links for $\pi_{1i,k}$ and $\pi_{2i,k}(t_1)$, so that $g_1^{-1}(\cdot) = g_2^{-1}(\cdot) = \mbox{expit}(\cdot)$. For the cross-sectional odds ratio, $\theta_{i,k}$, we used a log link, so that $g_\theta^{-1}(\cdot) = \mbox{exp}(\cdot)$. Based on these, we considered unstructured and B-spline specifications for the three sets of baseline parameters, $\BFalpha_1$, $\BFalpha_2$ and $\BFalpha_\theta$. For the latter, we used 10 knots for $\BFtau^{2.5}$ and 5 knots for $\BFtau^{5}$, with a cubic spline and a second order difference penalty (i.e. $m=2$ in $\Delta^m$), and considered varying degrees of penalty, specifically setting $\lambda_1$ = $\lambda_2$ = $\lambda_\theta$ = $\lambda$, with $\lambda \in \{0.0, 0.1, 0.5, 1.0, 2.5, 5.0\}$. Table A.17 in the Supplementary Materials reports AIC from the fitted models, across the different $\lambda$.

Finally, $\overline{\BFX}_{i1,k}$, $\overline{\BFX}_{i2,k}$, and $\overline{\BFX}_{i\theta,k}$ were specified with the overarching goal of assessing the joint impact of having $\ge$ 1 APOE $\epsilon$4 allele and gender on the joint risk of AD and death. As such, the models for $\pi_{1i,k}$ and $\theta_{i,k}$ included main effects for having $\ge$ 1 APOE $\epsilon$4 allele and gender, and their interaction, while the model for $\pi_{2i,k}(\cdot)$ included main effects, two-way interactions and the three-way interaction between AD diagnosis, having $\ge$ 1 APOE $\epsilon$4 allele and gender. For $\overline{\BFX}_{i1,k}$, $\overline{\BFX}_{i2,k}$, we additionally included all other variables available in dataset (i.e. race/ethnicity, marital status, education and depression).

\subsubsection{Baseline time trends}

\begin{figure}[h!]
\begin{subfigure}[t]{0.5\textwidth}
	\includegraphics[width=1\textwidth]{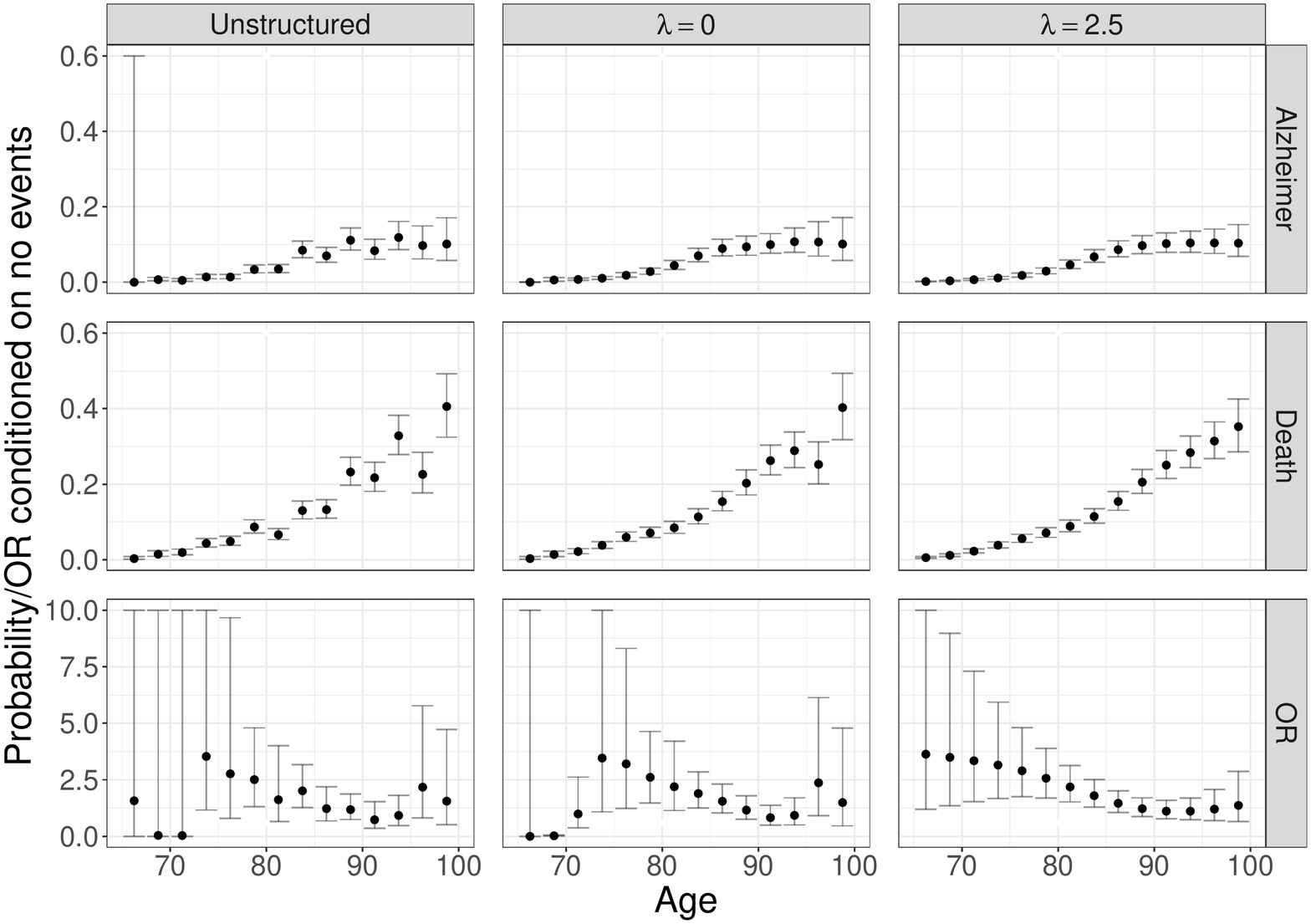}
	\caption{\footnotesize $\expit(\widehat{\BFalpha}_1)$, $\expit(\widehat{\BFalpha}_2)$ and $\exp(\widehat{\BFalpha}_\theta)$ under $\BFtau^{2.5}$.}
\end{subfigure}%
~~
\begin{subfigure}[t]{0.5\textwidth}
	\includegraphics[width=1\textwidth]{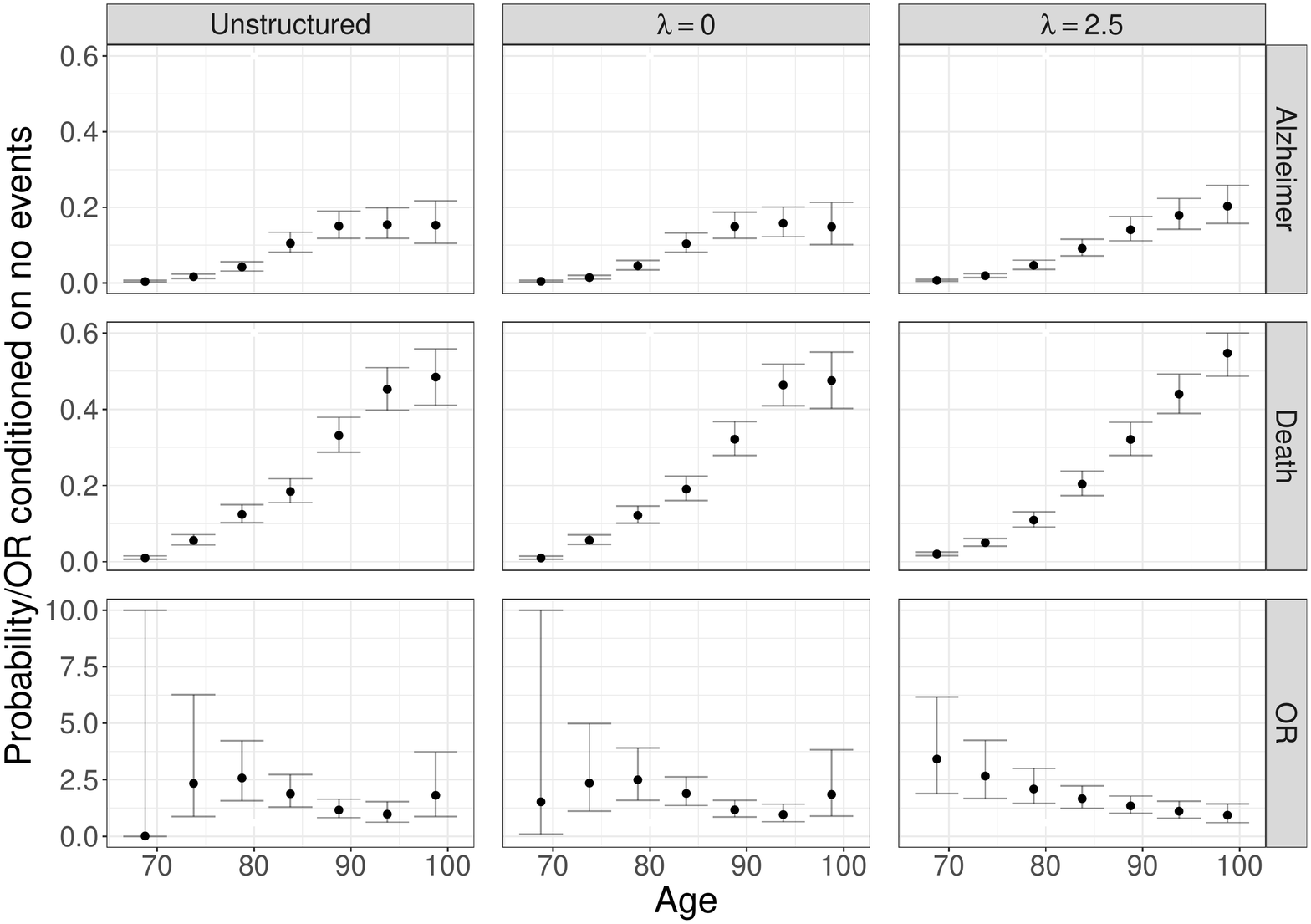}
	\caption{\footnotesize $\expit(\widehat{\BFalpha}_1)$, $\expit(\widehat{\BFalpha}_2)$ and $\exp(\widehat{\BFalpha}_\theta)$ under $\BFtau^{5.0}$.}
\end{subfigure}
\caption{Estimated baseline time trends, $\expit(\widehat{\BFalpha}_1)$, $\expit(\widehat{\BFalpha}_2)$ and $\exp(\widehat{\BFalpha}_\theta)$  from a series of analyses to the ACT data, under the unstructured specification and the B-spline specification with $\lambda \in \{0.0, 2.5\}$ and under the two partitions $\BFtau^{2.5}$ and $\BFtau^{5.0}$. See Section \ref{Sec:ACTanalysis:proposed} for details. Also shown are 95\% confidence intervals. Note, the y-axis has been truncated at 0.6 for $\expit(\widehat{\BFalpha}_1)$ and at 10 for $\exp(\widehat{\BFalpha}_\theta)$. For results under more $\lambda$ values, see Supplementary Figures A.7 and A.8.}
\label{Fig:ACTtimeTrends}
\end{figure}

Figure \ref{Fig:ACTtimeTrends} reports estimated baseline time trends, that is $\expit(\widehat{\BFalpha}_1)$, $\expit(\widehat{\BFalpha}_2)$ and $\exp(\widehat{\BFalpha}_\theta)$, under the unstructured and B-spline specifications. Note, given the coding of the variables included in the models, the interpretation of these quantities is specific to a population of individuals who are cognitively intact at age 65 and have the following characteristics: male, non-white, non-college educated, married without depression and no APOE $\epsilon$4 alleles.

Within each sub-figure, the estimates exhibit general concordance across the panels that differ in terms of structure and smoothing. Not surprisingly, the unstructured models exhibit greater uncertainty, especially in information-poor parts of the age scale (i.e. early on when there are relatively few AD events). From Figure \ref{Fig:ACTtimeTrends}(a), under the B-spline analysis with $\lambda$=2.5, the estimated baseline probability of an AD diagnosis during a given 2.5 year age interval, conditional on being AD-free and alive at the start of the interval, increases from 0.002 in $(65.0, 67.5]$ to 0.10 in $(97.5, 100.0]$. From Figure \ref{Fig:ACTtimeTrends}(b), the same increasing pattern emerges under partition $\BFtau^{5.0}$ with $\lambda$=0.0, specifically from 0.004 in $(65.0, 70.0]$ to 0.15 in $(95.0, 100.0]$. Note, that this is expected since the intervals are longer and, correspondingly, the cumulative number of events higher. Also from Figure \ref{Fig:ACTtimeTrends}(a), we see that the estimated baseline probability of death conditional on being event free during a given 2.5 year age interval increases from 0.006 in $(65.0, 67.5]$ to 0.35 in $(97.5, 100.0]$. Again, the same general pattern is observed under $\BFtau^{5.0}$, with the probability increasing from 0.01 in $(65.0, 70.0]$ to 0.48 in $(95.0, 100.0]$.

\subsubsection{Covariate effects for AD}

Based on the AIC criterion, the optimal values of $\lambda$ were $\lambda^*$=2.5 and $\lambda^*$=0.0 for $\BFtau^{2.5}$ and $\BFtau^{5.0}$, respectively. Table \ref{tab:ACT:results} reports estimated covariate associations from these fits for main effects and interactions for gender, having $\ge$ 1 APOE $\epsilon$4 allele and AD diagnosis (as appropriate); complete results are given in Tables A.18-A.20 of the Supplementary Materials.

\begin{sidewaystable}
\scriptsize
\begin{center}
\begin{tabular}{lcccccccc}
\hline\hline
	& \multicolumn{2}{c}{\textbf{Alzheimer's Disease, $\pi_{1i,k}$}} && \multicolumn{2}{c}{\textbf{Death, $\pi_{2i,k}$}} && \multicolumn{2}{c}{\textbf{Cross-sectional Odds Ratio, $\theta_{i,k}$}}\\
	& \multicolumn{2}{c}{$\exp(\widehat{\bbeta}_1)$} && \multicolumn{2}{c}{$\exp(\widehat{\bbeta}_2)$} && \multicolumn{2}{c}{$\exp(\widehat{\bbeta}_\theta)$}  \\
\cline{2-3} \cline{5-6} \cline{8-9}
	& $\BFtau^{2.5}$ & $\BFtau^{5.0}$ && $\BFtau^{2.5}$ & $\BFtau^{5.0}$ && $\BFtau^{2.5}$ & $\BFtau^{5.0}$ \\
\hline
\multicolumn{6}{l}{\textbf{Unstructured}} \\
~~Female 					& 0.97 (0.82, 1.15) & 0.99 (0.84, 1.18)	&& 0.57 (0.51, 0.64) & 0.55 (0.49, 0.63) && 0.80 (0.52, 1.22) & 0.96 (0.67, 1.39) \\
~~APOE$^a$					& 1.84 (1.46, 2.32) & 1.83 (1.44, 2.33)	&& 0.97 (0.80, 1.17) & 0.98 (0.81, 1.19) && 0.60 (0.30, 1.20) & 0.78 (0.44, 1.36) \\ 
~~Female $\times$ APOE$^a$ 		& 1.02 (0.76, 1.37) & 1.03 (0.76, 1.39)	&& 1.10 (0.85, 1.41) & 1.13 (0.87, 1.45) && 1.21 (0.49, 2.96) & 1.02 (0.50, 2.10) \\
~~AD$^b$					& & 								&& 2.71 (2.15, 3.41) & 2.81 (2.06, 3.82) \\
~~AD$^b$$\times$ Female  		& &								&& 1.57 (1.17, 2.09) & 1.83 (1.24, 2.71) \\
~~AD$^b$$\times$ APOE$^a$		& &								&& 1.70 (1.11, 2.59) & 2.37 (1.30, 4.29) \\
~~AD$^b$$\times$ Female $\times$APOE$^a$ & &						&& 0.61 (0.36, 1.03) & 0.48 (0.23, 1.01) \\
\multicolumn{6}{l}{\textbf{B-spline$^c$}} \\
~~Female 					& 0.97 (0.82, 1.14) & 0.99 (0.84, 1.17) 	&& 0.57 (0.51, 0.64) & 0.55 (0.49, 0.63) && 0.78 (0.51, 1.19) & 0.97 (0.67, 1.39) \\
~~APOE$^a$					& 1.83 (1.45, 2.32) & 1.83 (1.43, 2.34) 	&& 0.97 (0.80, 1.17) & 0.98 (0.81, 1.19) && 0.58 (0.29, 1.13) & 0.78 (0.44, 1.38) \\
~~Female $\times$ APOE$^a$ 		& 1.03 (0.77, 1.38) & 1.03 (0.76, 1.40) 	&& 1.10 (0.85, 1.41) & 1.13 (0.87, 1.46) && 1.20 (0.50, 2.92) & 1.02 (0.49, 2.11) \\
~~AD$^b$					& & 							 	&& 2.64 (2.08, 3.34) & 2.81 (2.04, 3.88) \\
~~AD$^b$$\times$ Female  		& &							 	&& 1.56 (1.16, 2.09) & 1.82 (1.22, 2.74) \\
~~AD$^b$$\times$ APOE$^a$		& &							 	&& 1.70 (1.10, 2.64) & 2.37 (1.25, 4.49) \\
~~AD$^b$$\times$ Female $\times$APOE$^a$ & &				 	 	&& 0.62 (0.36, 1.06) & 0.48 (0.22, 1.04) \\
\hline\hline
\multicolumn{6}{l}{\scriptsize $^a$ An indicator of having at least one APOE $\epsilon$4 allele.} \\
\multicolumn{6}{l}{\scriptsize $^b$ An indicator of whether the patient has previously had a diagnosis of Alzheimer's disease} \\
\multicolumn{6}{l}{\scriptsize $^c$ Based on the optimal $\lambda$ of $\lambda^*$=2.5 for $\BFtau^{2.5}$ and $\lambda^*$=0.0 for $\BFtau^{5.0}$.}
\end{tabular}
\end{center}	
\caption{\label{tab:ACT:results}Select results, specific to the role of gender and having $\ge$ 1 APOE $\epsilon$4 allele on the joint risk of AD and death, based on the proposed framework applied to the data from the Adult Changes in Thought study. Complete results are given in Tables A.18-A.20 of the Supplementary Materials.}
\end{sidewaystable}

From the first two columns of Table \ref{tab:ACT:results}, the results regarding risk of AD are largely consistent across the four sets of analyses based on the two partitions and on use of either an unstructured or B-spline specification for the baseline time trends. Specifically, we find that, while having $\ge$ 1 APOE $\epsilon$4 allele is a strong risk factor, there is no evidence that gender is a risk factor nor that it is an effect modifier of the APOE effect; both point estimates are close to the null.

\subsubsection{Covariate effects for death and global dependence}

The middle set of results in Table \ref{tab:ACT:results} speak to how gender, having $\ge$ 1 APOE $\epsilon$4 allele and having a diagnosis of AD jointly influence risk of mortality. Note the four components involving a diagnosis of AD correspond to $\BFbeta_{2,y}$ in model (\ref{Eq:p2:model}) and, thus, jointly represent global dependence between AD and mortality. As with the results regarding the risk of AD, the results for death are largely consistent between the unstructured and baseline specifications. Interestingly, there are some differences between those based on the $\BFtau^{2.5}$ partition and those based on the $\BFtau^{5.0}$ partition although the overarching conclusion that there is an important interplay between the three factors in determining risk of mortality is consistent between the two. To facilitate discussion of the results, Table \ref{tab:ACT:ORdeath} reports odds ratio estimates and 95\% confidence intervals for mortality, based on the B-spline specification models reported in Table \ref{tab:ACT:results}, across combinations of whether the patient has had an AD diagnosis, their gender, and whether they have $\ge$ 1 APOE $\epsilon$4 allele. From the first four rows, in the absence of a diagnosis of AD, female gender is associated with approximately 40\% lower odds of mortality while there is no evidence to indicate that the number of APOE $\epsilon$4 alleles play a role.

\begin{table}[h!]
\footnotesize
\begin{center}
\begin{tabular}{ccccc}
\hline\hline
AD		& Gender		& \# APOE		& \multicolumn{2}{c}{Odds ratio (95\% CI)} \\
\cline{4-5}
status	& 			& $\epsilon$4 alleles & $\BFtau^{2.5}$ & $\BFtau^{5.0}$ \\
\hline
No AD	& Male		& 0		& 1.00			& 1.00 \\
No AD	& Male		& $\ge$ 1	& 0.97 (0.80, 1.17)	& 0.98 (0.81, 1.19) \\
No AD	& Female		& 0		& 0.57 (0.51, 0.64)	& 0.55 (0.48, 0.62) \\
No AD	& Female		& $\ge$ 1	& 0.61 (0.51, 0.72)	& 0.61 (0.51, 0.72) \\
AD		& Male		& 0		& 2.64 (2.08, 3.34)	& 2.81 (2.04, 3.88) \\
AD		& Male		& $\ge$ 1	& 4.35 (3.09, 6.14)	& 6.53 (3.83, 11.1) \\
AD		& Female		& 0		& 2.35 (1.94, 2.84)	& 2.81 (2.17, 3.65) \\
AD		& Female		& $\ge$ 1	& 2.64 (2.09, 3.34)	& 3.54 (2.54, 4.94) \\
\hline\hline
\end{tabular}
\end{center}	
\caption{\label{tab:ACT:ORdeath}Odds ratio estimates and 95\% confidence intervals for mortality across combinations of whether the patient has had an AD diagnosis, their gender, and whether they have $\ge$ 1 APOE $\epsilon$4 allele. Results are based on the B-spline models reported in Table \ref{tab:ACT:results}.}
\end{table}

From the lower half of Table \ref{tab:ACT:ORdeath}, the odds of mortality among those patients with a diagnosis of AD diagnosis, and, thus, extent of global dependence, is substantially higher with the magnitude depending on the interplay between gender and the number of APOE $\epsilon$4 alleles. Moreover, while having at least APOE $\epsilon$4 alleles seems to be associated with higher odds, the increase is substantially larger for males.  

\subsubsection{Local dependence}

Turning to the assessment of local dependence, the third row in the two sub-figures of Figure \ref{Fig:ACTtimeTrends} suggest that there are meaningful time trends in the co-occurrence of AD and death, specifically that the risk of co-occurrence within a given interval are highest at the early ages. From the B-spline specification with $\lambda$=2.5 applied to partition $\BFtau^{2.5}$, for example, the odds ratio among males with no APOE $\epsilon$4 alleles decreases from 3.63 (95\% CI: 1.19, 11.04) during the $(65.0, 67.5]$ interval to 1.37 (95\% CI: 0.66, 2.86) during the $(97.5, 100.0]$. From Table \ref{tab:ACT:results}, although the confidence intervals all include 1.00, the point estimates for all three covariates in the model applied to the $\BFtau^{2.5}$ partition are indicative of clinically meaningful associations. For example, the local dependence odds ratio is estimated to be 42\% smaller among males with $\ge$ 1 APOE $\epsilon$4 allele compared to those without. The corresponds odds ratio for females is estimated to be approximately 30\% smaller (0.58 $\times$ 1.20 $\approx$ 0.70) for those with $\ge$ 1 APOE $\epsilon$4 allele compared to those without.

Finally, comparing the local dependence results between those based on $\BFtau^{2.5}$ and those based on $\BFtau^{5.0}$ indicate that the choice of partition can have a meaningful impact on the conclusions. This is also seen in the results regarding global dependence (see Tables \ref{tab:ACT:results} and \ref{tab:ACT:ORdeath}). Moreover, under $\BFtau^{5.0}$ the evidence regarding an interaction between gender and having $\ge$ 1 APOE $\epsilon$4 allele on local dependence is  weaker when the partition intervals are 5 years in length.


\section{Discussion}
\label{Sec:Discussion}

Although less familiar than competing risks, semi-competing risks arise in a wide range of clinical settings including: Alzheimer's disease, as illustrated in this paper; graft-versus-host disease among patients who have undergone hematopoietic stem cell transplantation for leukemia~\citep{ferrara2009}; readmission following discharge from a hospitalization during which a patient is diagnosed of a terminal disease such as pancreatic cancer~\citep{lee2015bayesian, lee2016hierarchical}; shock among patients with implanted cardiac devices~\citep{reeder2019joint}; and, preeclampsia a severe pregnancy-associated disease that affects between 3-10\% of all pregnancies~\citep{jeyabalan2013epidemiology}. A distinguishing feature of semi-competing risks, relative to competing risks, is that there is at least partial information about the joint distribution between $T_1$ and $T_2$. In this paper, we propose a new methodology that seeks to directly leverage this information in order to gain interpretable insight into dependence between the two events. Because the interpretation of the model components, including the notions of global and local dependence, are distinct from the interpretations one obtains with existing methods, we view the proposed framework as being complimentary to existing methods (as opposed to seeking to replace them). Moreover, we view the proposed framework as being in-line with recent work that seeks to better understand whether and how specific factors confer risk jointly on multiple outcomes, such as the \textit{dual hazard rate} proposed by Prentice and Zhao \cite{prentice2020regression}.

The foundation for the proposed framework is the discretization of the time scale. As discussed in Sections \ref{SubSec:ProposedFramework:joint} and \ref{Sec:ProposedFramework:choice}, one cannot say that a given choice of $\BFtau$ is the `truth' and yet the specific choice dictates the numerical values and interpretation of the results. This results in a tension that is illustrated in Table \ref{tab:ACT:results}: one cannot claim that either $\BFtau^{2.5}$ or $\BFtau^{5.0}$ is the `right' choice and yet there are instances where the numerical results differ in meaningful ways. Our view of this dilemma is that consideration of multiple partitions can be viewed as an opportunity to obtain additional insight. Consider, for example, the main effect of female gender in the model for $\theta_{i,k}$ based on the B-spline specification: under $\BFtau^{2.5}$ the estimated impact of female gender is to reduce the local dependence odds ratio by 22\% while the reduction is only 3\% under $\BFtau^{5.0}$. Thus, ignoring the lack of statistical significance, for the purpose of discussion, there is an indication that among individuals with no APOE $\epsilon$4 alleles, gender plays a role in the co-occurrence of AD and death over relatively short time frames (i.e. 2.5 years) but not over longer time frames (i.e. 5 years). Conceptually this is analogous to the results that one might see in a Cox model for a univariate outcome if the effect of a covariate varies over time (i.e. non-proportional hazards) and yet proportional hazards is adopted; in such settings, the value of the common hazard ratio that is being estimated will depend on the interval over which data is available.

A number of extensions to the proposed framework are possible. From a theoretical perspective, an interesting alternative to having $\blambda$ fixed is to consider $\blambda=\blambda_{N}=o(1)$; that is the amount of regularization decreases as more data become available. This framing of the asymptotics, however, while resulting in consistent estimation if the true functions coincide with a B-spline, leads to a variance expression that do not involve $\blambda$ and thus has been perceived as of less useful in practice \citep{gray1992flexible, yu2002penalized}. Second, as mentioned in Section \ref{Sec:ACT} follow-up in ACT consists of biennial visits. As such, while the date of death can be precisely ascertained through death records, AD is subject to interval censoring. Generalizing our approach to interval censored data will likely be an interesting challenge, therefore, since some event times cannot be straightforwardly assigned to a specific interval under given a partition. Furthermore, a related problem may arise in defining covariate values when time-dependent covariates are observed only intermittently \cite{nevo2020novel}. 

\bibliographystyle{agsm}
\bibliography{LongitSemiComp}
\end{document}